**Paper title**

Unlocking capacities of viral genomics for the COVID-19 pandemic response

**Authors and Affiliations**

† These authors contributed equally to the work.

* These authors contributed equally to the work.

Sergey Knyazev†

Department of Computer Science, College of Art and Science, Georgia State University, 1 Park Place, Room 618, Atlanta, GA 30303, USA

sergey.n.knyazev@gmail.com

ORCID: https://orcid.org/0000-0003-0385-1831

Twitter: SeKnyaz

Karishma Chhugani†

Department of Pharmacology and Pharmaceutical Sciences, School of Pharmacy, University of Southern California, 1985 Zonal Avenue, Room 713. Los Angeles, CA 90089, USA

chhugani@usc.edu

ORCID: https://orcid.org/0000-0003-1328-5028

Varuni Sarwal*

Department of Computer Science, University of California Los Angeles, 580 Portola Plaza, Los Angeles, CA 90095, USA

sarwal8@gmail.com

ORCID: https://orcid.org/0000-0001-7563-9835

Ram Ayyala*

Department of Neuroscience, College of Life Sciences, University of California Los Angeles, 580 Portola Plaza, Los Angeles, CA 90095, USA

ramayyala@gmail.com

ORCID:https://orcid.org/0000-0001-7275-271X

Harman Singh*

Department of Electrical Engineering, Indian Institute of Technology, Hauz Khas, New Delhi, 110016, India

harmansingh.iitd@gmail.com

ORCID:https://orcid.org/0000-0002-3970-6276

Smruthi Karthikeyan

Department of Pediatrics, University of California, San Diego, La Jolla, CA, USA

skarthikeyan@health.ucsd.edu

Dhrithi Deshpande

Department of Pharmacology and Pharmaceutical Sciences, School of Pharmacy, University of Southern California, 1985 Zonal Avenue, Room 713. Los Angeles, CA 90089, USA

ddeshpan@usc.edu

ORCID: https://orcid.org/0000-0003-3794-7364

Zoia Comarova

Paradigm Environmental, 3911 Old Lee Highway, Fairfax, VA 22030

zoia.comarova@gmail.com

Angela Lu

Department of Pharmacology and Pharmaceutical Sciences, School of Pharmacy, University of Southern California, 1985 Zonal Avenue, Room 713. Los Angeles, CA 90089-9121, USA

alu52904@usc.edu

Yuri Porozov

World-Class Research Center "Digital biodesign and personalized healthcare", I.M. Sechenov First Moscow State Medical University, Moscow, Russia

Department of Computational Biology, Sirius University of Science and Technology, Sochi, Russia

yuri.porozov@gmail.com

Aiping Wu

Center for Systems Medicine, Institute of Basic Medical Sciences, Chinese Academy of Medical Sciences & Peking Union Medical College, Beijing, 100005, China

Suzhou Institute of Systems Medicine, Suzhou, 215123, China

wap@ism.cams.cn

Malak S. Abedalthagafi

Genomics Research Department, Saudi Human Genome Project, King Fahad Medical City and King Abdulaziz City for Science and Technology, Riyadh, Saudi Arabia

malthagafi@kacst.edu.sa

ORCID: https://orcid.org/0000-0003-1786-3366

Twitter: @malakabed

Shivashankar H. Nagaraj

Centre for Genomics and Personalised Health, Queensland University of Technology, Brisbane, QLD 4059, Australia

Translational Research Institute, Brisbane, Australia

shivashank@gmail.com

Adam L. Smith

Astani Department of Civil and Environmental Engineering, University of Southern California, 3620 South Vermont Avenue, Los Angeles, CA 90089

smithada@usc.edu

Pavel Skums

Department of Computer Science, College of Art and Science, Georgia State University, 1 Park Place, Floor 6, Atlanta, GA 30303, USA

pskums@gsu.edu

Jason Ladner

The Pathogen and Microbiome Institute, Northern Arizona University, Flagstaff, AZ 86011

jason.ladner@nau.edu

Tommy Tsan-Yuk Lam

State Key Laboratory of Emerging Infectious Diseases, School of Public Health, The University of Hong Kong

ttylam@hku.hk

Nicholas C. Wu

Department of Biochemistry, University of Illinois at Urbana-Champaign, Urbana, IL 61801, USA

Carl R. Woese Institute for Genomic Biology, University of Illinois at Urbana-Champaign, Urbana, IL 61801, USA

Center for Biophysics and Quantitative Biology, University of Illinois at Urbana-Champaign, Urbana, IL 61801, USA

nicwu@illinois.edu

ORCID: http://orcid.org/0000-0002-9078-6697

Alex Zelikovsky

Department of Computer Science, College of Art and Science, Georgia State University, 1 Park Place, Floor 6, Atlanta, GA 30303, USA

The Laboratory of Bioinformatics, I.M. Sechenov First Moscow State Medical University, Moscow, 119991, Russia

alexz@gsu.edu

ORCID:https://orcid.org/0000-0003-4424-4691

Rob Knight

Department of Pediatrics, University of California, San Diego, La Jolla, CA, USA

Department of Bioengineering, University of California, San Diego, La Jolla, CA, USA

Department of Computer Science & Engineering, University of California, San Diego, La Jolla, CA, USA

Center for Microbiome Innovation, University of California, San Diego, La Jolla, CA, USA

robknight@eng.ucsd.edu

Keith A. Crandall

Computational Biology Institute and Department of Biostatistics & Bioinformatics, Milken Institute School of Public Health, George Washington University, Washington, DC 20052

kcrandall@gwu.edu

Serghei Mangul

Department of Clinical Pharmacy, School of Pharmacy, University of Southern California, 1540 Alcazar Street, Los Angeles, CA 90033, USA

serghei.mangul@gmail.com

ORCID: https://orcid.org/0000-0003-4770-3443

Twitter: smangul1

**Abstract**

More than any other infectious disease epidemic, the COVID-19 pandemic has been characterized by the generation of large volumes of viral genomic data at an incredible pace due to recent advances in high-throughput sequencing technologies, the rapid global spread of SARS-CoV-2, and its persistent threat to public health. However, distinguishing the most epidemiologically relevant information encoded in these vast amounts of data requires substantial effort across the research and public health communities. Studies of SARS-CoV-2 genomes have been critical in tracking the spread of variants and understanding its epidemic dynamics, and may prove crucial for controlling future epidemics and alleviating significant public health burdens. Together, genomic data and bioinformatics methods enable broad-scale investigations of the spread of SARS-CoV-2 at the local, national, and global scales and allow researchers the ability to efficiently track the emergence of novel variants, reconstruct epidemic dynamics, and provide important insights into drug and vaccine development and disease control. Here, we discuss the tremendous opportunities that genomics offers to unlock the effective use of SARS-CoV-2 genomic data for efficient public health surveillance and guiding timely responses to COVID-19.

## Introduction

COVID-19, a contagious disease caused by the novel severe acute respiratory syndrome coronavirus 2 (SARS-CoV-2), has reached an extraordinary scale not seen since the influenza pandemic of 1918–1919[1]. Within a month of its first reported case in China in December 2019, COVID-19 had spread to many regions in China[2] and had been detected in several neighboring countries, including Thailand, Korea, and Japan. As international flights continued to operate, SARS-CoV-2 then spread to Europe and North America in a short amount of time, and was soon declared a global pandemic[3,4]. According to the World Health Organization (WHO), in the first 16 months of the pandemic (through April 8, 2021), more than 132.7 million people became infected worldwide, resulting in more than 2.8 million deaths[5].

Over the past two decades, the biomedical community has become equipped with infrastructure for basic genomic techniques to support epidemic responses[6], a capability that has enabled the rapid collection of SARS-CoV-2 genomic information which has allowed observation of SARS-CoV-2 genomic evolution online, rapid tracking of SARS-CoV-2 genetic groups, lineages, variants, variants of interest (VOI) and variants of concern (VOC)[7]. The precise and rapid tracking of SARS-CoV-2 genetic changes facilitates fast development of SARS-CoV-2 clinical tests and predicting the efficiency of the vaccines. As sequencing technologies and genomic analysis tools progress, genome sequencing is becoming more widely integrated into clinical and healthcare workflows. However, the utilization of genomic sequencing to its full potential for public health surveillance and outbreak response efforts has yet to be established and depends on the broad expansion of the best practices for preventing and limiting outbreaks that had been determined during the COVID-19 response[8]. Herein, we discuss the genomic capacities that can be used to address many of the  public health issues associated with COVID-19 (**Box 1**).

**Initial SARS-CoV-2 detection and characterization**

Genomic analysis conducted on respiratory specimens isolated from the first COVID-19 patients hospitalized in December, 2019, in Wuhan, China, allowed for the prompt detection and characterization of a novel coronavirus, later named SARS-CoV-2, by January 2020[3,4,9,10]. Initial sequence analyses revealed that SARS-CoV-2 shared 80% nucleotide identity with SARS-CoV[11,12], strongly indicating that SARS-CoV-2 was likely a respiratory pathogen that could spread from human to human and hence with clear epidemic potential. These initial analyses also revealed that SARS-CoV-2 shared high sequence similarity with related viruses found in bats and pangolins, suggesting a zoonotic origin[10,11,13–19]. Across its complete genome, SARS-CoV-2 is most closely related to the bat coronavirus RaTG13, with which it shares approximately 96% nucleotide sequence identity. However, different SARS-CoV-2 coding regions share greater similarity to those of other animal coronaviruses. For example, the spike (S) protein receptor-binding domain (RBD) exhibits higher sequence identity (97.4%) to that of the Guangdong pangolin virus, rather than to RaTG13 (89.3%), while SARS-CoV-2 long 1ab (replicase) open reading frame (ORF) exhibits the highest sequence identity (98.8%) with the RmYN02 bat coronavirus[20]. In further support of zoonotic origin, another coronavirus detected in five bats (RacCS203) is genetically closely related to SARS-CoV-2[21], and neutralizing antibodies for SARS-CoV-2 were found in wild pangolins and bats from Thailand[21]. Moreover, there is similarity between SARS-CoV-2 zoonosis and the zoonoses of the SARS-CoV and MERS-CoV coronaviruses, as data indicate that in all three cases, other intermediate animals were likely present in their transmission chains[22,23]. Together, these findings suggest a complex history of recombination events prior to the zoonotic transfer of SARS-CoV-2 to humans, although when and in which hosts these events took place remains unclear[13,14,19,24]. Genomic knowledge acquired through viral sequencing and phylogenetic analysis greatly contributed to the rapid determination of the potential epidemiological characteristics and origins of SARS-CoV-2[25].

**The role of genomics in the early COVID-19 outbreak response**

As seen with other recent viral epidemics, viral genome sequencing has become an essential part of the COVID-19 public health response[26]. Early access to SARS-CoV-2 genome sequences allowed for the timely development and production of nucleic acid amplification testing (NAAT)-based diagnostics, expedited vaccine development, and accelerated opportunities for SARS-CoV-2 genomics-based real-time surveillance[27–32]. The first SARS-CoV-2 tests and vaccine candidates appeared within one and three months, respectively, after the identification of the first COVID-19 patient[32–34]. Together with access to modern sequencing technologies , the scale of the pandemic, based on numbers of cases and affected regions, has prompted the collection of SARS-CoV-2 viral genomic data at an unparalleled magnitude (on average 2,500 genomes per day). Consequently, the capacity to track virus spread and evolution in real time has been accelerated relative to that associated with prior outbreaks[35]. When the WHO initially declared a Public Health Emergency of International Concern (PHEIC) on January 30, 2020, 339 SARS-CoV-2 genomes had already been collected and characterized[2–4,28]. By April 7, 2021, public repositories that host SARS-CoV-2 genomes contained over 1,000,000 genomes[36–38] (**Table S1**). Notably, by the end of the sixth month of the pandemic (May 2020), Global Initiative on Sharing All Influenza Data (GISAID) and the National Center for Biotechnology Information (NCBI) databases included 110,000 SARS-CoV-2 full-length genome sequences as compared to more than the 8,000 HIV full-length genome sequences collected by the Los Alamos sequence National Laboratory[39] over the past 40 years[40] (**Figure 1a**). 86% of available SARS-CoV-2 raw sequencing data at NCBI is Illumina data, 13.7% is Oxford Nanopore, and 0.3% is Pacbio, IonTorrent and BGISEQ (**Figure S1**). There is a correlation between the number of submitted sequences per capita and the GDP per capita for the majority of the countries in the world, moreover, high-income countries submitted about 100x more sequences per capita on average than did low-income countries (**Figure 1b, S2**). However, it is remarkable that African nations with a low GDP per capita sequenced viral genomes on a level comparable to that of middle-and high-income countries[41]. Indeed, due to several previous programs that were aimed at controlling outbreaks of other viruses in Africa, the sequencing capacity of the African healthcare system improved, helping to increase its efficiency in the sequencing of SARS-CoV-2 genomes[41] (**Figure 1c**). Countries with the highest ratios for numbers of SARS-CoV-2 genomes sequenced to numbers of COVID-19 cases and relatively low number of reported cases per capita were Taiwan, New Zealand, Australia, Iceland, and Denmark (**Figure 1d**).

**SARS-CoV-2 genomic evolution**

The unprecedented scale of SARS-CoV-2 genome sequencing offers unique opportunities for tracking SARS-CoV-2 evolution online and detecting the emergence and spread of new VOI and VOC[42–44] (**Figure 2**). Due to SARS-CoV-2 genome sequencing and consequent bioinformatics analysis it was shown that because of an intrinsic RNA proofreading mechanism, coronaviruses exhibit lower mutation rates than do many other RNA viruses, such as Ebola virus and HIV[45–48]. In addition, their evolutionary (i.e., nucleotide substitution) rate partly reflects the action of host-dependent RNA-editing enzymes (e.g., APOBEC)[49]. Coronaviruses undergo a mean rate of approximately $1.12 \times 10^{-3}$ nucleotide substitutions per site per year. This is comparable to the SARS-CoV-1 mutation rate from $0.8 \times 10^{-3}$ to $2.38 \times 10^{-3}$, Ebolavirus's mutation rate of $1.3 \times 10^{-3}$ and is lower than seasonal influenza mutation rate of $6.7 \times 10^{-3}$ and HIV mutation rate of $4.4 \times 10^{-3}$ [45–47,50–52].

Another important aspect of SARS-CoV-2 evolution is that SARS-CoV-2, like many other RNA viruses, can live in the host as a swarm of closely related variants within individual hosts and has a tendency for recombinations[53]. Genomic studies have demonstrated the presence of such intra-host diversity inside hosts[54–60], with one study having identified between 1 and 52 haplotype variants in each of 25 clinical patients[54]. Identifying the factors that shape these intra-host viral population structures can promote a better understanding of short-term viral evolution, in addition to providing insights into host adaptation and drug and vaccine design. For example, evidence of intra-host recombination[61] may enable estimating the role of recombination in the zoonotic origin of SARS-CoV-2[14] and the emergence of novel viral variants[62–64].

Over the first year of the epidemic, SARS-CoV-2 has gradually accumulated mutations and developed into several viral lineages as it has spread through the human population[7,65–68]. However, from the advent of the pandemic through approximately September 2020, there was no statistical evidence that any of the numerous characterized SARS-CoV-2 mutations had resulted in a loss or gain of function[45–47]. For example, one study analyzed all 48,454 SARS-CoV-2 genomes available from GISAID from late July of 2020 that had been sequenced throughout the world and identified 12,706 mutations, 398 of which were recurrent, and none of which were associated with a significant change in transmissibility[69]. During the summer of 2020, the D614G mutation in the viral S protein sparked attention because this new variant globally superseded the original SARS-CoV-2 strain globally. Phylogenetic analyses and clinical evidence indicated that, although the D614G variant was associated with both increased viral load and infectivity[68,70], it was also more susceptible to neutralizing antisera and was not linked to any change in vaccine efficacy or increased pathogenicity[71].

The first SARS-CoV-2 viral variant of concern (VOC) for public health, known as variant B.1.1.7, was first detected in the UK in September 2020. Genomic analysis revealed that this B.1.1.7 variant had first arisen in late Summer or early Fall 2020, and then quickly spread through many countries, including Australia, Denmark, Italy, Iceland, the Netherlands, and now the US[72–74]. However, the full pathogenic potential of this variant was not recognized until December 2020[72]. The B.1.1.7 variant strain harbors at least 12 mutations, including 2 in the S protein: N501Y, which increases the ability of SARS-CoV-2 binding to its cellular receptor, ACE2, and P618H, which adjoins the furin cleavage site in the S protein[75–77]. Both mutations have been associated with a 40-80% increase in the transmissibility of this variant as compared to previous SARS-CoV-2 strains[72] . More recently, the B.1.1.7 variant was found to be associated with greater disease severity and an increased risk of death as compared to other variants[78]. In addition, the variant carries a Δ69–70 deletion that results in detection failure by some SARS-CoV-2 molecular tests, which can limit the successful tracing of this VOC[79]. However, there is no evidence thus far that this variant reduces vaccine efficacy.

The second VOC was discovered in, UK, in September 2020, and was characterized by several mutations, including E484K in the RBD of the S protein. This mutation, which was later discovered to have arisen independently in other viral variants around the world[80], is associated with reduced neutralizing activity of human convalescent and post-vaccination sera. Additional VOCs related to B.1.1.7 include B.1.351, which was first detected in South Africa in November 2020,[81] where it spread rapidly. Although the latest reports indicate that this variant has also spread to Zambia and the US, there is no evidence that this mutation impacts disease severity[82]. This variant also harbors multiple mutations in the S protein, such as K417N, E484K, and N501Y.

The third VOC, P.1, was detected in four travelers who arrived in Japan from Brazil in January 2021[83–85]. P.1 carries similar mutations in the RBD domain as B.1.351 (K417T, E484K, N501Y), the latter of which can increase transmissibility and help the virus evade neutralizing antibodies. The impact of the K417T mutation is not known. More recently, another genetic variant B.1.427/B.1.429 was declared as VOC because of its prevalence in the outbreak that happened in California. This variant is harboring the L452R mutation in the S protein that is suspected to confer SARS-CoV-2 antibody resistance, although it is less severe than the E484K mutation, which is associated with greatly reduced viral susceptibility to antibody neutralization[86]. The full and actual list of all VOI and VOC can be found at the official CDC page[87].

Some of these variants were first independently identified in immunodeficient individuals in different countries, suggesting that their emergence may be the result of convergent evolution followed by rapid spread. For example, the appearance of the ΔH69/ΔV70 deletion was documented in an immunosuppressed individual through deep viral genome sequencing at 23 time points during the course of infection (101 days)[64]. A weakened host immune response can permit the virus to replicate with little or no control, increasing the likelihood for mutations to occur. The independent evolution of a given mutation in different geographic locations suggests that this mutation may confer an adaptive advantage to the virus, such as immune evasion or increased transmissibility, which is corroborated by clinical studies. Given the likely public health importance of these VOCs and VUIs, global surveillance for these and other new variants is expanding, as information for all SARS-CoV-2 lineages is now collected and made available online for the rapid evaluation of their epidemiologic and vaccine impact and short-term evolution based on individual data points[7,88]. In order to gain better control over emerging VOC and VOI, A European Commission Recommendation dated 19 January 2021 stated that "all EU Member States should reach a capacity of sequencing at least 5% - and preferably 10% - of positive test results. In most Member States, the sequencing capacity for identification of SARS-CoV-2 variants is below the recommendation set by the European Commission to sequence 5-10% of SARS-CoV-2 positive specimens'' (**Figure 1d**).

**The use of genomics to investigate the pandemic spread of SARS-CoV-2**

Access to rich and diverse publicly available SARS-CoV-2 genomic data across various regions has allowed scientists and public health officials to efficiently track routes along which COVID-19 outbreaks have spread locally and internationally (**Figure S3**). In this context, phylogenetic and genetic network analyses can provide important public health information regarding viral epidemic spread[89,90]. Importantly, as viruses accumulate genomic mutations within different populations, knowledge regarding such evolution can reveal transmission chains and distinguish imported cases from instances of local transmission if a sufficient number of samples is analyzed, ultimately identifying high risk transmission routes which should be subject to enhance public health control[91–94]. Genomic analysis has allowed the identification of SARS-CoV-2 introduction into Europe from China, into the US from both China and Europe with a subsequent local transmission[91,95–97]. One recent study suggests that SARS-CoV-2 was introduced in the US in Connecticut via a domestic transmission route, while another showed that most successful viral introductions to Arizona were likely from domestic travel[91,98]. Another study revealed that the New York City area exhibited multiple introductions of SARS-CoV-2, primarily from Europe[99]. Similarly, SARS-CoV-2 was potentially introduced into France from several countries, including China, Italy, the United Arab Emirates, Egypt, and Madagascar[92]. We have curated a comprehensive list of genomic outbreak investigations to date for various geographical regions (**Table S2**). This catalog contains 40 studies and is updated in real-time as more studies are published; an online version is available at https://github.com/Mangul-Lab-USC/COVID-19-outbreak-investigations.

Viral genomics can also be used to monitor the effectiveness of global travel restrictions and lockdowns in different countries in limiting viral spread. For example, genomic analysis showed that the risk of domestic transmission of SARS-CoV-2 in Connecticut exceeded that of international introduction at the time federal travel restrictions were imposed, highlighting the critical need for local surveillance[91]. Similarly in Brazil, three clades of European origin were established prior to the initiation of travel bans and lockdowns[100]. Another genomic analysis showed that, due to violations of imposed lockdowns with sea trade, several SARS-CoV-2 international introductions likely occurred in Morocco[101]. In Australia, lockdown effectiveness was validated using agent-based modeling coupled with SARS-CoV-2 genomic data[102]. On December 19, 2019, due to the new rapidly spreading B.1.1.7 variant found in the UK, the prime minister implemented tighter lockdown and other restrictions, and as a result, many countries closed their borders to people traveling from the UK[103]. The spread of this variant then was precisely tracked in the U.S. due to available sequencing data[104].

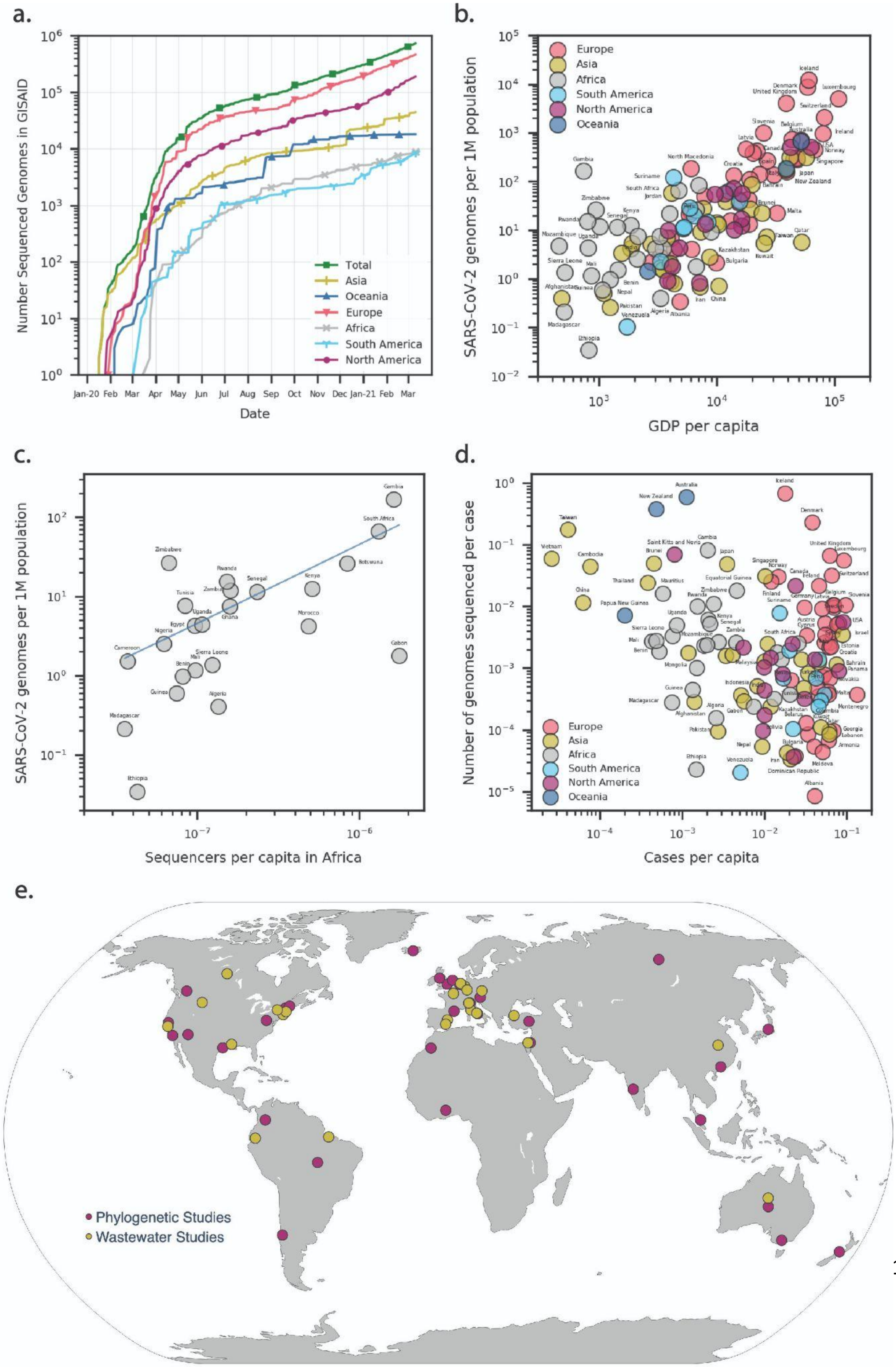

a.
Number Sequenced Genomes in GISAID
Date
Jan-20 Feb Mar Apr May Jun Jul Aug Sep Oct Nov Dec Jan-21 Feb Mar
Total
Asia
Oceania
Europe
Africa
South America
North America
b.
SARS-CoV-2 genomes per 1M population
GDP per capita
Europe
Asia
Africa
South America
North America
Oceania
c.
SARS-CoV-2 genomes per 1M population
Sequencers per capita in Africa
d.
Number of genomes sequenced per case
Cases per capita
Europe
Asia
Africa
South America
North America
Oceania
e.
Phylogenetic Studies
Wastewater Studies

**Figure 1. Available SARS-CoV-2 genomic sequencing data and its usage for outbreak investigation** (a) The number of global SARS-CoV-2 genomes sequenced according to Global Initiative On Sharing All Influenza Data (GISAID) between January 2020-March 2020. (b) The number of available SARS-CoV-2 sequences in GISAID per 1 million (1M) individuals for each country vs. the number of cases per capita up to March 2021. (c) The number of available SARS-CoV-2 sequences in GISAID per 1 million (1M) individuals for each country in Africa vs. the number of sequencers per capita up to March 2021. Blue line is a correlation line of all data points on the plot (d) The number of available SARS-CoV-2 sequences in GISAID per number of reported COVID-19 cases for each country vs. the number of reported COVID-19 cases per capita up to March 2021. (e) Global outbreak investigations by phylogenetic analysis (red) and wastewater studies (yellow), dots were placed in the geographical centers of each county or region.

Combining genomic methods with clinical and geospatial data can help characterize viral infectivity, virulence, and death rates of circulating viral strains more accurately because epidemics in different areas of the world may have distinct characteristics that depend on viral genotype, as well the demographics of the host population. Specifically, integration and analysis of phylogenetic and epidemiologic data can provide a more complete understanding of the pandemic transmission dynamics[105]. Available genomic data can also be utilized to examine and partly explain the relationship between genetic variation in strains of SARS-CoV-2 and disease severity[106]. Findings from these studies can also help characterize mutation patterns in various hotspots and identify correlates of infection and death rates in these countries[107]. Novel approaches can be developed to combine population genomics and genetics to leverage the identification of molecular markers with unusual pattern variations or relevant single nucleotide polymorphisms in people from different geographies[67,68]. If SARS-CoV-2 infections continue at their current rate, population genomic research and pharmacogenomics approaches may be useful in the development of personalized therapeutics against this pathogen. Although disease severity can be partly attributed to host genomics, understanding these factors has been difficult due to contradictory evidence and limited host genomics studies conducted thus far to date[108].

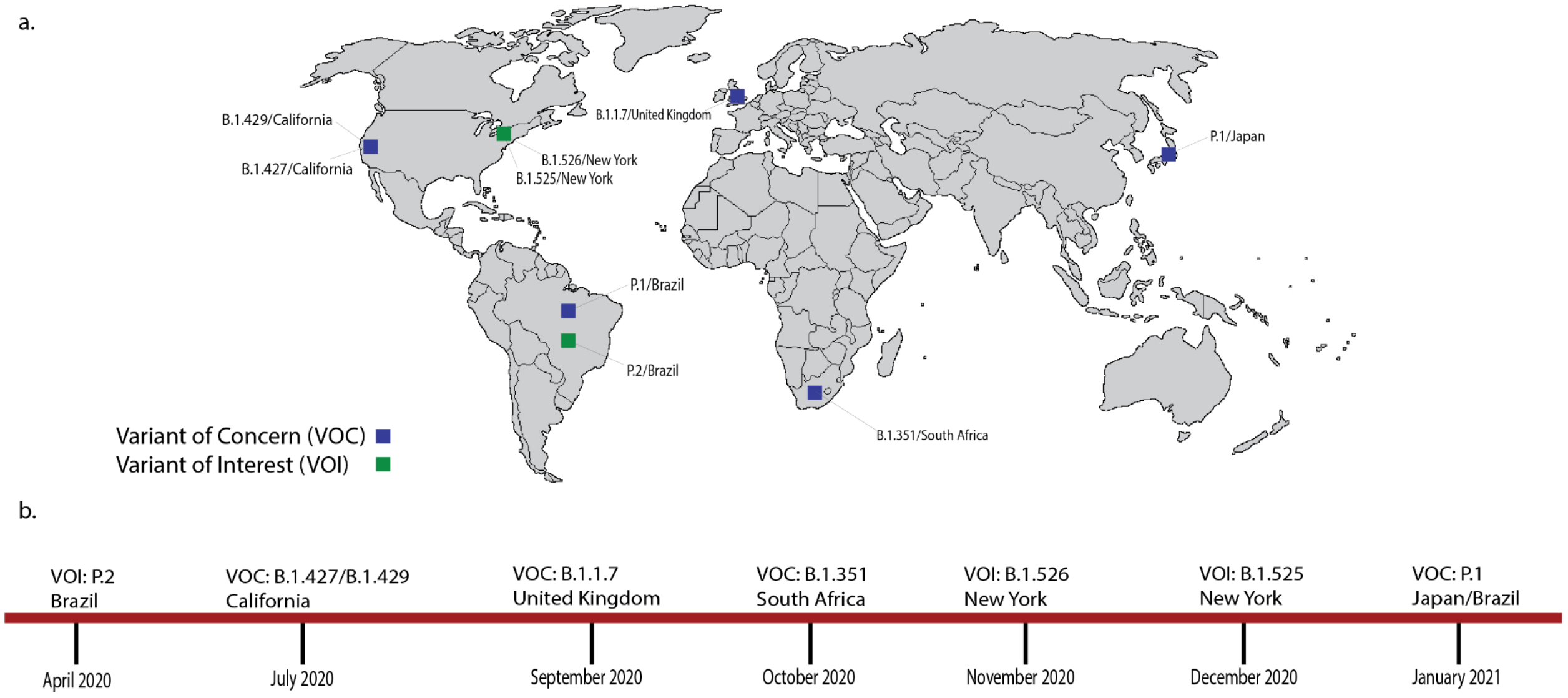

**Figure 2. Variant of Concern (VOC) and Variant of Interest (VOI) circulating throughout the globe** (a) The locations where VOCs and VOIs were initially detected. (b) The timeline showing when VOCs and VOIs initially appeared in the sequencing data (not the time when they were declared as VOCs and VOIs).

## Monitoring SARS-CoV-2 transmission through wastewater genomic studies

Another genomics-based method for population-level pathogen surveillance assesses the presence of trace viral genomic material in wastewater, with this approach having been successfully employed to track antibiotic use[109] and tobacco consumption[110] and for the monitoring of enteric viruses such as poliovirus[111]. Notably, a 2013 study accomplished the early detection of a viral outbreak in Sweden by quantifying hepatitis A virus and norovirus genetic material levels in wastewater[112]. Although COVID-19 is primarily associated with respiratory symptoms, SARS-CoV-2 is regularly shed in feces[113]. As of August 2020, SARS-CoV-2 RNA had been detected in wastewater by over 35 studies in 17 countries using NAAT-based methods (https://www.covid19wbec.org), which can effectively detect the presence and concentration of viruses in wastewater[114] and potentially estimate the relative number of disease cases in the area covered by the sewage facilities. However, current NAAT-based methods cannot detect whether or not these samples harbor novel mutations[115], and the development of novel mutations in the template primer binding sites have the potential to compromise the efficacy of NAAT-based methods to detect the viral presence[116]. Additionally, wastewater may contain fragmented or defective genomes which may not be detected with these methods. Alternatively, a potentially promising approach is the application of metagenomics on a global scale to detect, collect, and store samples in preparation for future pandemics[117,118]. Metagenomics methods, which can sequence all available genomic material in a sample, allow the characterization of an entire viral population and the detection of prevalent SARS-CoV-2 variants in a given geographical space[119,120]. Wastewater surveillance studies of SARS-CoV-2 RNA concentration across various regions in the world have taken place between January 2020 and November 2020. (**Figure 1e, Table S3**).

Temporal changes in SARS-CoV-2 RNA concentration in wastewater were assessed in Valencia, Spain from February till April 2020, Paris, France from March till April 2020, and in many other regions (see **Table S3**), they have been consistent with the number of clinically diagnosed cases in a given community[121,122]. This relationship demonstrates the use of wastewater studies as a relatively inexpensive and straightforward method for investigating national outbreak dynamics, especially in areas where case diagnosis is complicated. In contrast, clinical diagnostic testing traditionally used to assess the number of cases in a community typically underestimates actual infection rates[123], as this approach primarily focuses on symptomatic individuals because the asymptomatic cases are less likely to be captured. However, combining clinical diagnostics with wastewater-based surveillance can potentially provide a more comprehensive community-level profile of both symptomatic and asymptomatic cases, enabling identification of hospital capacity needs[114,124–130]. Additionally, an important advantage of wastewater monitoring is the ability to detect early-stage outbreaks before they become widespread[111,115,131,132]. In contrast to NAAT-based methods such as real time reverse transcription polymerase chain reaction (RT-PCR)-based analysis of SARS-CoV-2, metagenomic sequencing allows for characterization of the prevalent SARS-CoV-2 genomic variants in a defined local region and reveal geospatial SARS-CoV-2 genotype distribution[120,133]. Using wastewater samples can identify circulating lineages in the community and accompany analysis of genomic epidemiology, for example, such analysis has already helped to detect B.1.1.7 strains in the US and Switzerland[134].

Despite the numerous advantages of wastewater-based virus surveillance, many potential improvements would result in more reliable and extended applications in public health decision-making. Currently, wastewater-based methods require calibration and validation because they only provide a raw measure of the number of cases in a population[115]. Additionally, wastewater-based monitoring lacks the granularity of clinical diagnostic testing and cannot discern a particular area of an outbreak when the wastewater treatment plant serves a large population. Sampling at a higher spatial resolution within the sewer system or even at a building-level scale could potentially provide early indications of viral outbreaks and help monitor their progression[135]. This effort could also include areas with large numbers of septic tank systems that are not feeding municipal wastewater systems.

**Genomics in clinical applications**

Viral genomics can also aid in vaccine development and investigations of how viral evolution impacts clinical outcomes and treatment[136]. While the majority of known SARS-CoV-2 mutations have no effect on viral replication and transmission[137], some substitutions have been linked to phenotypic changes that could influence outbreak dynamics. For example, patients in Singapore infected with $\Delta$382 SARS-CoV-2 variants, which have a 382-nucleotide deletion in ORF8, exhibited milder symptoms compared to patients with viruses that lacked this deletion[138]. However, the $\Delta$382 SARS-CoV-2 variant is very rare globally and appears to have died out in Singapore. More alarmingly, the highly prevalent D614G mutation may increase transmissibility and infectivity in natural populations, giving variants harboring this mutation a marked selective advantage, although the best evidence to date comes from laboratory and simulation studies only[68,95,139]. Lastly, the viral variants B.1.1.7, B.1.351, and P.1, which were detected in late 2020, showed significantly increased transmissibility, heightening concern from a public health perspective[82,140,141]. For vaccine development, understanding the degree to which different regions of the viral genome are prone to mutation is important, as it is necessary to understand whether rising immunity in humans will result in antigenic drift and consequent vaccine escape. These evolutionary effects are commonly seen, for example, in human influenza viruses and endemic coronaviruses. Analyses of the current genomic variability of SARS-CoV-2 suggest that prospective COVID-19 vaccines should be cross-protective for the majority of currently known viral variants[73,108,142], although some minor variants (< 1% natural occurrence frequency) have been shown to alter the antigenicity of SARS-CoV-2[143]. For vaccine development, determining the structures of SARS-CoV-2 antigens and their mutants is also crucial for the maximization of vaccine efficacy[144]. The online COVID-3D resource allows for the exploration of the structural distribution of genetic variation in SARS-CoV-2[145].

The antigenic drift may also affect the effectiveness of NAAT-testing because when mutations happen in primer regions, the effectiveness of tests drops due to loss of affinity. Therefore control over appearing mutations should be taken into account for updating NAAT tests as well.

The first lab-confirmed case of COVID-19 re-infection case was detected in Hong Kong using a genomic analysis approach[146], after which additional re-infection cases were detected in Belgium, Ecuador, and the US[147–149]. Phylogenetic analyses of longitudinal SARS-CoV-2 genomic sequences for all these patients distinguished between patient re-infection and persistent viral shedding from the initial infection. The findings in all four cases suggest that SARS-CoV-2 may persist in the global human population, despite herd immunity due to natural infection, which can complicate vaccine development and efficacy[146]. However, current data suggests that SARS-CoV-2 re-infection is rare, and it has been proposed that immunity against reinfection can last for at least several months after the primary infection[150].

Clinical manifestations of SARS-CoV-2 infection vary greatly, ranging from a lack of symptoms to irreversible pulmonary damage[151–153]. Adaptive immune responses, such as early CD8+ and CD4+ T cell responses, have been associated with positive patient outcomes[154]. Next-generation sequencing of T and B cell receptor repertoires from COVID-19 patients has also revealed differences in immune response characteristics between patients with a mild or severe disease course[155]. Schultheiß et al. detected more than 14 million T and B receptors from blood samples of infected patients from 70 time points, compiling  a valuable resource that can inform new therapeutic approaches and vaccine development. For example, their study revealed that knowledge of host immunopathology obtained through sequencing can permit the early detection of clinical biomarkers and aid in the identification of patients at risk for severe disease[155].

**Integrating clinical and genomics data**

Many SARS-CoV-2 genomic studies to date have been conducted in the absence of substantial clinical data collection and/or integration with viral sequence data. Conversely, numerous studies have critically evaluated extensive clinical data alone, without assessing corresponding genomic data[156]. Even investigations yielding large genomic (e.g., GISAID[28,36]) and clinical datasets[157] have not performed integrated analyses of *both* data types. This substantial limitation of current practices results from distinctions between the fields of bioinformatics (genomic data analyses) and medical informatics (clinical data analyses). The COVID-19 pandemic promises to unite researchers from both of these fields to integrate these seemingly disparate data sources, especially in prospective studies[158]. Finding significant associations between genomic and clinical features of the virus will ultimately support more targeted interventions by public health officials.

**Discussion**

The unprecedented density and volume of available SARS-CoV-2 genomic and clinical data enabled the prompt and effective characterization of both SARS-CoV-2 genomes and COVID-19 epidemiology compared to those of previous outbreaks. The numerous successful efforts across various parts of the globe utilizing genomic data for addressing the COVID-19 outbreak created a solid foundation for the standardization of using SARS-CoV-2 genomic data. High-income countries sequenced more SARS-CoV-2 sequences per population than the countries with low, middle-low and middle income. However, the countries of Africa with low and middle-low income demonstrated remarkably better preparedness to collect SARS-CoV-2 genomes than low and middle-low income countries from other continents (**Figure 1b**). This preparedness can be attributed to previous global initiatives to support African countries in mitigating previous outbreaks of other viruses that ended up in growing sequencing capacity of the region. Africa provides remarkable examples of the necessity of international cooperation which should be implemented in other parts of the globe for better control of worldwide epidemiology.

At the same time, the unprecedented volume of SARS-CoV-2 genome sequencing that reached one million viral genomes sequences challenged the current practices of viral data storage, processing, and bioinformatics analysis[159–161]. While the importance of genome-based viral surveillance systems was widely recognized, the principle of such systems were conceptualized, and there were technological burdens of creating them, as such systems were still in the early stages of development before the pandemic started. However, the unprecedented mobilization of financial, scientific, and development resources during the course of COVID-19 allowed for fast development, deployment, and scaling of numerous global surveillance systems which provide resources for outbreak response using SARS-CoV-2 genome analysis (**Table 1**).

When rigorously studied, benchmarked, and standardized, viral genomic surveillance systems enable reliable and timely detection of the presence of circulating and emerging pathogens similar to SARS-CoV-2, providing us a robust shield from current and newly emerging outbreaks[162]. With sufficient sampling, genomic analysis will enable sentinel surveillance efforts capable of effectively locating the geographic source of outbreaks, elucidating transmission chains, and ultimately limiting the spread of the pathogens globally[99,141,163–167].

**Acknowledgements**

We thank William M. Switzer and Ellsworth M. Campbell from the Division of HIV/AIDS Prevention, Centers for Disease Control and Prevention, Atlanta, 30333 GA, USA for useful discussions and suggestions. We also thank numerous anonymous reviewers who helped improve our manuscript by their valuable comments on the manuscript.

**Funding**

S.M. was partially supported by National Science Foundation grants 2041984. Tommy Lam is supported by NSFC Excellent Young Scientists Fund (Hong Kong and Macau) (31922087) and Health and Medical Research Fund (COVID190223). Pavel Skums was supported by the National Institutes of Health grant 1R01EB025022. Malak Abedalthagafi MA a acknowledge King Abdulaziz City for Science and Technology and the Saudi Human Genome Project for technical and financial support (https://shgp.kacst.edu.sa) Nicholas Wu: startup funds from the University of Illinois at Urbana-Champaign Adam Smith: acknowledge funding from NSF grant no. 2029025. Alex Zelikovsky: A.Z. has been partially supported by NSF Grant CCF-1619110 and NIH Grant 1R01EB025022-01. Sergey Knyazev S.K. has been partly supported by Molecular Basis of Disease at Georgia State University. Rob Knight: NSF project 2038509, RAPID: Improving QIIME 2 and UniFrac for Viruses to Respond to COVID-19. CDC project 30055281 with Scripps led by Kristian Andersen, Genomic sequencing of SARS-CoV-2 to investigate local and cross-border emergence and spread.

**Box 1: The value of capacities of genomic analysis for COVID-19 pandemic responses**

1. **Initial SARS-CoV-2 virus detection and characterization.** Genomic analysis conducted on respiratory specimens isolated from the first COVID-19 patients hospitalized in December, 2019, in Wuhan, China, allowed for the prompt detection and characterization of a novel coronavirus, later named SARS-CoV-2, by January 2020[3,4,9,10]. Initial sequence analyses revealed that SARS-CoV-2 shared 80% nucleotide identity with SARS-CoV[11,12], strongly indicating that SARS-CoV-2 was likely a respiratory pathogen that could spread from human to human and hence with clear epidemic potential. These initial analyses also revealed that SARS-CoV-2 shared high sequence similarity with related viruses found in bats and pangolins, suggesting a zoonotic origin[10,11,13–19]. Across its complete genome, SARS-CoV-2 is most closely related to the bat coronavirus RaTG13, with which it shares approximately 96% nucleotide sequence identity. However, different SARS-CoV-2 coding regions share greater similarity to those of other animal coronaviruses.
2. **The role of genomics in the early COVID-19 outbreak response.** Early access to SARS-CoV-2 genome sequences allowed for the timely development and production of nucleic acid amplification testing (NAAT)-based diagnostics, expedited vaccine development, and accelerated opportunities for SARS-CoV-2 genomics-based, real-time surveillance[27–32]. By April 7, 2021, public repositories that host SARS-CoV-2 genomes contained over 1,000,000 genomes[36–38] (**Table S1**). Notably, by the end of the sixth month of the pandemic (May 2020), Global Initiative on Sharing All Influenza Data (GISAID) and the National Center for Biotechnology Information (NCBI) databases included 110,000 SARS-CoV-2 full-length genome sequences as compared to more than the 8,000 HIV full-length genome sequences collected by the Los Alamos sequence National Laboratory[39] over the past 40 years[40] (**Figure 1a**).

3. **SARS-CoV-2 genomic evolution.** The unprecedented scale of SARS-CoV-2 genome sequencing offers unique opportunities for tracking SARS-CoV-2 evolution online and detecting the emergence and spread of new variants of interest (VOI) and variants of concern VOC[42–44] (**Figure 2**). Coronaviruses undergo a mean rate of approximately 1.12 x $10^{-3}$ nucleotide substitutions per site per year. This is comparable to the SARS-CoV-1 mutation rate from 0.8 x $10^{-3}$ to 2.38 x $10^{-3}$, Ebolavirus's mutation rate of 1.3 x $10^{-3}$ and is lower than seasonal influenza mutation rate of 6.7 x $10^{-3}$ and HIV mutation rate of 4.4 x $10^{-3}$ [45–47,50–52]. However, during the first year and a half of SARS-CoV-2 evolution there 3 VOI and 5 VOC appeared in the world.

4. **The use of genomics to investigate the pandemic spread of SARS-CoV-2.** Online tracking of changes in SARS-CoV-2 genomes allows for detecting viral transmission routes. Genomic analysis has allowed the identification of SARS-CoV-2 introduction into Europe from China, into the US from both China and Europe with a subsequent local transmission[91,95–97]. We have collected around 40 outbreak investigations available in the literature that illustrates the usability of genomic analysis for tracking pathogen spread routes.

5. **Monitoring SARS-CoV-2 transmission through wastewater genomic studies.** SARS-CoV-2 is regularly shed in feces[113] that allows surveying the COVID-19 epidemiology by testing wastewater samples. That allows estimating the number of COVID-19 cases, the prevalence of the lineages in the region, and detecting new viral variants. We have collected over 35 studies in 17 countries that used wastewater studies for detecting the presence and concentration of viruses in wastewater[114] and estimate the relative number of disease cases in the area covered by the sewage facilities.

6. **Genomics in clinical applications.** Viral genomics aids in detecting how viral evolution impacts clinical outcomes and treatment[136]. For example, SARS-CoV-2 lineages that have been linked to phenotypic changes that could influence outbreak dynamics and clinical outcomes were declared as VOIs and VOCs. Genomics also helps to control over antigenic drift that can influence vaccines and test efficiency. This provides an opportunity to proactively adapt vaccines and tests to ongoing SARS-CoV-2 genome changes.

7. **Integrating clinical and genomics viral genomics data.** Many SARS-CoV-2 genomic studies to date have been conducted in the absence of substantial clinical data collection. Conversely, numerous studies have critically evaluated extensive clinical data alone, without assessing corresponding genomic data[156]. This substantial limitation of current practices results from distinctions between the fields of bioinformatics (genomic data analyses) and medical informatics (clinical data analyses). The COVID-19 pandemic promises to unite researchers from both of these fields to integrate these seemingly disparate data sources, especially in prospective studies[158].

**End of Box 1**

**Table 1: Online services with SARS-CoV-2 genome resources and analytics**

| **Resource** | **Description** | **Link** |
|---|---|---|
| GISAID | Assembled genome database and analysis | https://www.gisaid.org/ |
| NCBI | Raw sequencing data database | https://www.ncbi.nlm.nih.gov/sars-cov-2/ |
| COG-UK | United Kingdom sequences database | https://www.cogconsortium.uk/ |
| PANGO | Lineage analytics | https://cov-lineages.org/ |
| Nextstrain | Phylogenetic analysis | https://nextstrain.org/ |
| WBEC | Wastewater analytics | https://www.covid19wbec.org/ |
| COVID-3D | Structural changes of lineages | http://biosig.unimelb.edu.au/covid3d/ |

**Supplementary Figures**

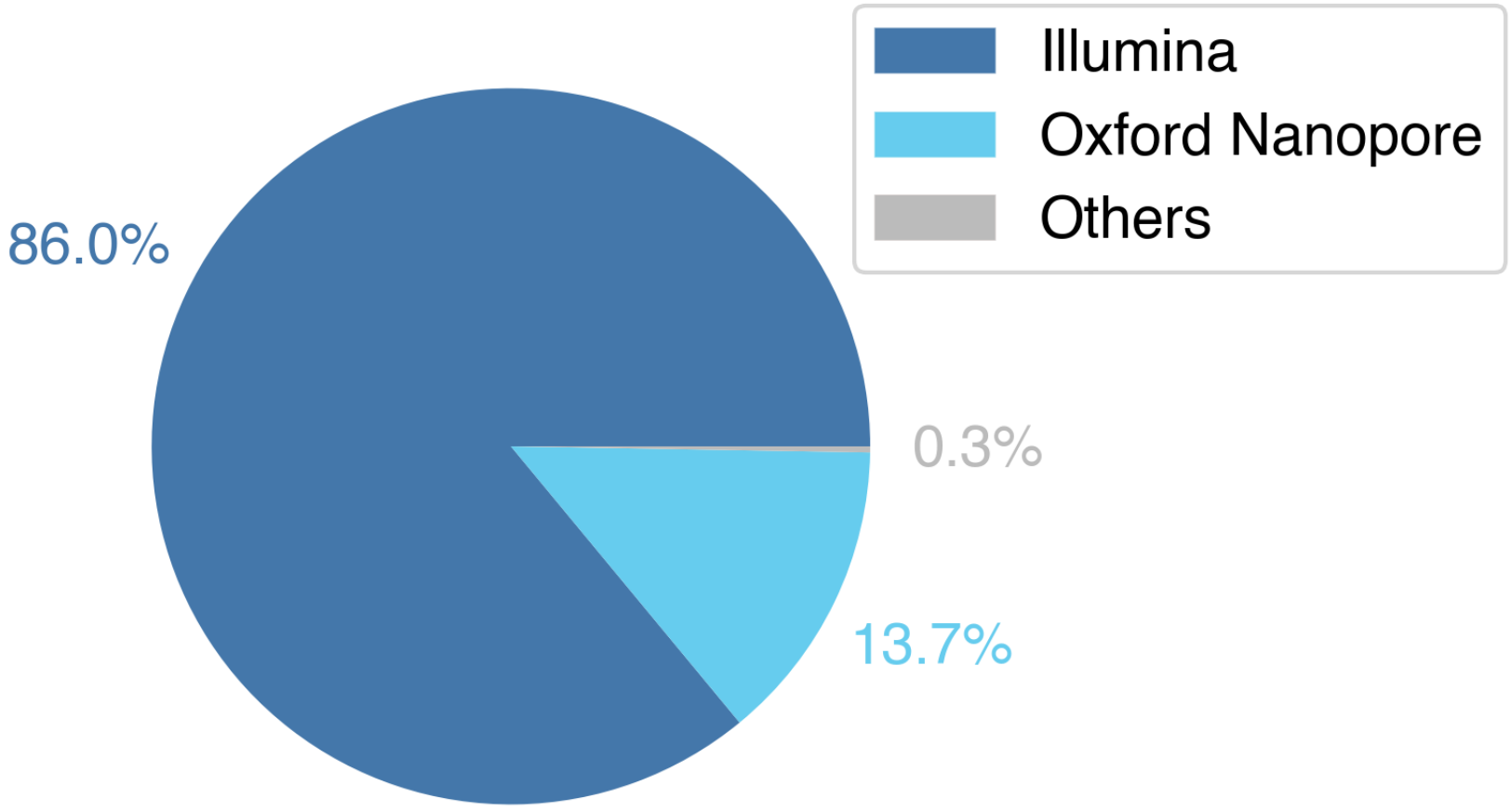

**Figure S1** Sequenced genomes present on SRA database sequenced by Illumina, Nanopore, and others sequencing technologies that include Pacbio, IonTorrent and BGISEQ.

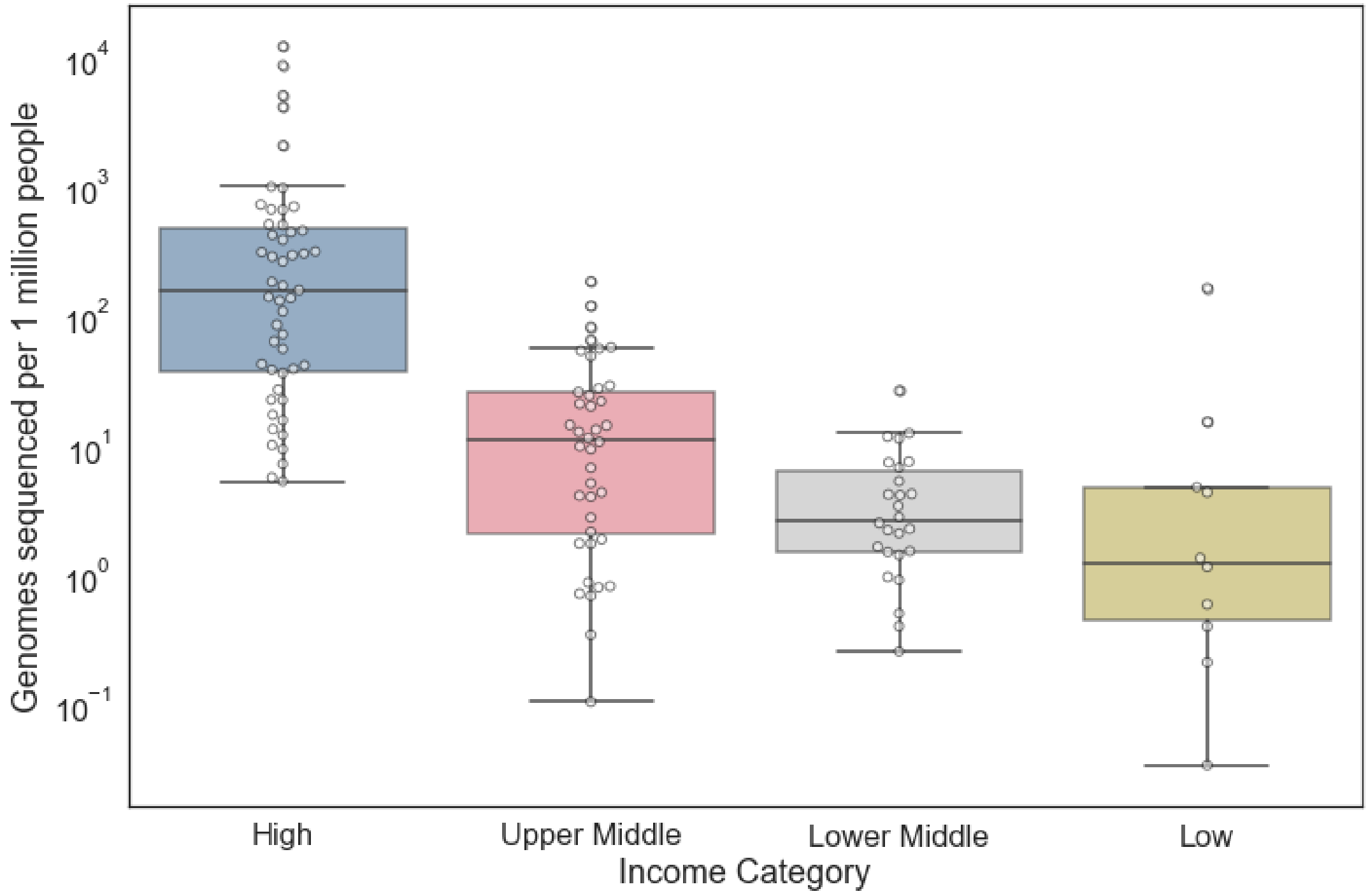

**Figure S2** The number of genomes sequenced per million people per income category

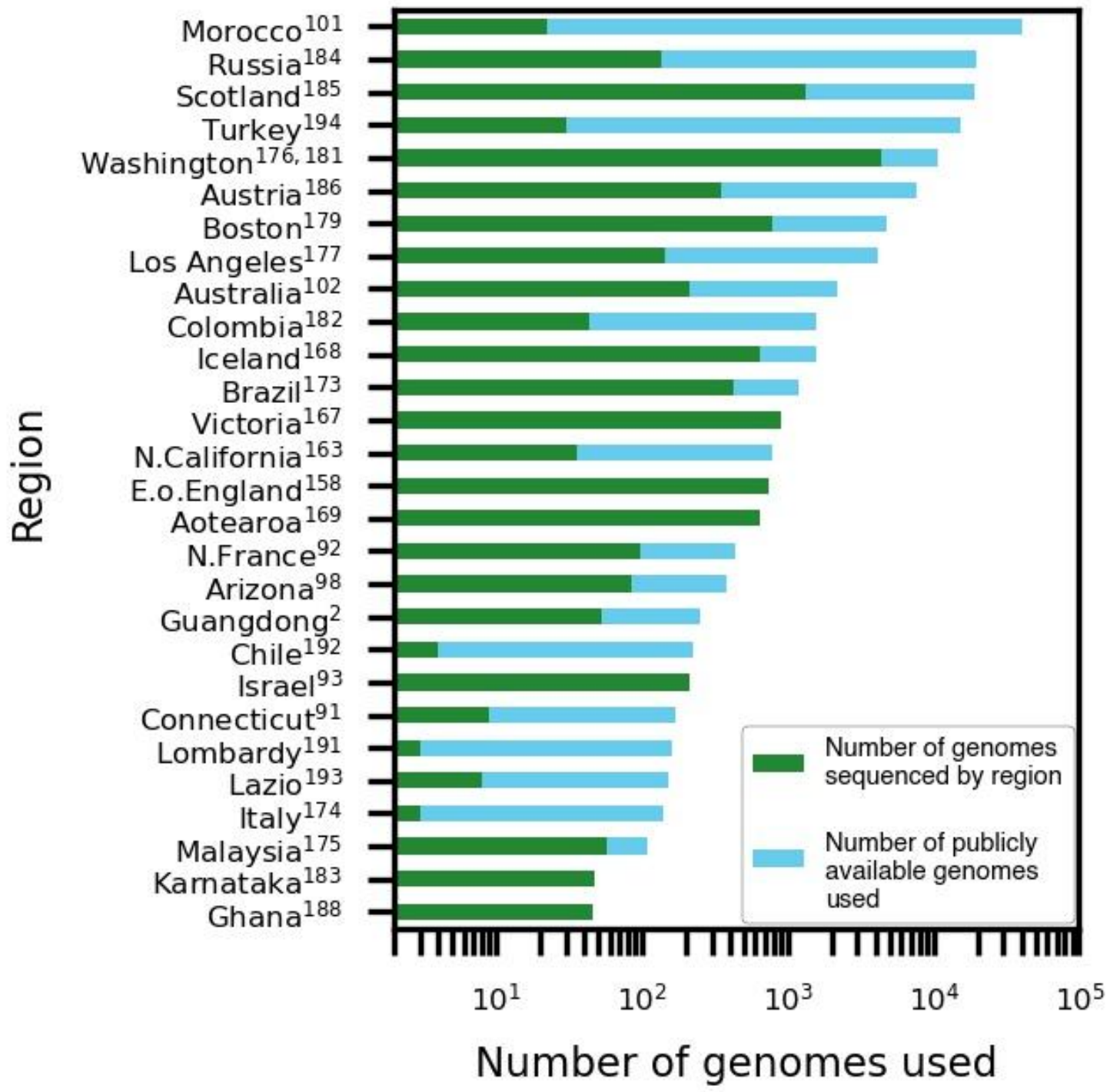

**Figure S3** Number of genomes sequenced and publicly available across regions that were used for outbreak investigation studies in different regions.

**Supplementary Table 1: Public repositories collecting SARS-CoV-2 genomes.**

| Database | Genomes* | Raw data* | Link |
| --- | --- | --- | --- |
| GISAID | 1,013,054 | - | https://www.gisaid.org/ |
| COG-UK | 413,687 | - | https://www.cogconsortium.uk/ |
| NCBI GenBank | 151,209 | 333,257 | https://www.ncbi.nlm.nih.gov/sars-cov-2/ |

* - Submitted by April 7, 2021

**Supplementary Table 2: Summary of genomic analyses of SARS-CoV-2**

| The region covered (country) | Date of publication | Date of sample collection | Number of genomes sequenced by country/region | Total number of genomes used for analysis (combined with publicly available data) | Phylogeny Method | Citation |
|---|---|---|---|---|---|---|
| Guangdong Province (China) | April 30 2020 | January 9-31, 2020; February 5, 2020, February 10-12, 2020 | 53 | 250 | -Maximum likelihood phylogeny | [2] |
| Northern France | April 29, 2020 | December 2019 | 97 | 438 | -Maximum likelihood phylogeny | [92] |
| Connecticut (United States) | May 28, 2020 | March 14, 2020 | 9 | 168 | -Maximum likelihood phylogeny<br>-Clade-defining nucleotide substitution | [91] |
| Victoria, Australia | September 1, 2020 | January 6, 2020 - April 14, 2020 | 903 | 903 | -Maximum likelihood phylogeny<br>-Bayesian maximum clade credibility tree<br>-Bayesian | [168] |
| | | | | | phylodynamic analysis | |
| Iceland | June 11, 2020 | March 31, 2020 - April 1, 2020 | 643 | 1547 | -Median-joining haplotype network | 169 |
| Israel | May 22, 2020 | March 17, 2020 - April 22, 2020 | 212 | 212 | -Maximum likelihood phylogeny<br>-Phylogeographic analysis | 93 |
| Northern California (United States) | July 31, 2020 | January 29, 2020 - March 20, 2020 | 36 | 789 | -Maximum likelihood phylogeny | 164 |
| East of England (United Kingdom) | July 14, 2020 | March 13, 2020 - April 25, 2020 | 747 | 747 | -Maximum likelihood phylogeny | 158 |
| Aotearoa (New Zealand) | August 20, 2020 | February 26, 2020 - May 22, 2020 | 649 | 649 | -Maximum likelihood time-scaled<br>-Bayesian maximum clade credibility tree | 170 |
| Multi-country<br>Australia<br>India<br>United Kingdom<br>China | August 10, 2020 | May 27, 2020 | <br>1364<br>109<br>400 | 18,168 | -Maximum likelihood phylogeny<br>-Haplotype networks | 171 |
| United States | | | 112<br>365 | | | |
| Netherlands | July 16, 2020 | February 25 -28, 2020; March 1 - 11, 2020 | 189 | Combined with all available full-length SARS-CoV-2 genomes from GISAID as of March 22, 2020 | -Maximum likelihood phylogeny<br>-Bayesian phylogeny<br>-Maximum clade credibility tree | 167 |
| Japan | June 5, 2020 | March 5 - 15, 2020 | 10 | Combined with all available full-length SARS-CoV-2 genomes from GISAID as of March 30, 2020 | -Haplotype networks | 172 |
| Morocco | June 25, 2020 | June 7, 2020 | 22 | 40,390 | -Maximum likelihood phylogeny<br>-Time resolved tree | 101 |
| Global | April 28, 2020 | March 4, 2020 | 160 | 413 | -Phylogenetic network analysis | 173 |
| Australia | July 9, 2020 | January 21, 2020 - March 28, 2020 | 209 | 2194 | -Genomic clustering<br>-Maximum likelihood phylogeny<br>-agent-based | 102 |
| | | | | | modeling | |
| Brazil | July 23, 2020 | March 5, 2020 - April 30, 2020 | 427 | 1182 | -Maximum likelihood and molecular clock phylogenies<br><br>-Time-resolved maximum clade credibility phylogeny | [174] |
| Italy | March 27, 2020 | N/A | 3<br><br>(Sequences from GISAID) | 141 | -Maximum likelihood phylogeny | [175] |
| Malaysia | August 27, 2020 | February - April 2020 | 58 | 108 | -Maximum likelihood phylogeny | [176] |
| Washington (US) | October 30, 2020 | February 20, 2020 - March 15, 2020 | 455 | 493 | -Maximum likelihood phylogeny | [177] |
| Los Angeles, California (US) | September 18, 2020 | February 28, 2020 - June 22, 2020 | 142 | 4095<br><br>(286<br><br>Sequences from LA county, of which 144 are obtained from | -Maximum parsimony phylogeny | [178] |
| | | | | GISAID<br>+ 3809<br>genomes from across the world) | | |
| Houston, Texas (US) | September 29, 2020 | March 6, 2020 - July 7, 2020 | 5,085 | Combined with genomes acquired through GISAID on 19 August 2020. | -Maximum likelihood phylogeny | [179] |
| Baltimore, Maryland (US) | August 23, 2020 | March 11 - 31, 2020 | 114 | Combined with datasets from GISAID having 886 and 2593 genomes | -Maximum likelihood phylogeny | [94] |
| Boston, Massachusetts (US) | August 25, 2020 | January 29, 2020 - April 18 2020<br>March 4, 2020 - May 9, 2020 | 772 | 4783 | Markov chain Monte Carlo Phylodynamics | [180] |
| Norfolk (United Kingdom) | September 30, 2020 | March 2020 - August 2020 | 1035 | The phylogenetic tree was estimated as part of the COG-UK phylogenetic pipeline (2020-09-07) | Maximum likelihood phylogeny | [181] |
| Washington (US) | September 30, 2020 | February 2020 - July 2020 | 3940 | 10051 | Markov chain Monte Carlo Phylodynamics | [182] |
| Arizona (US) | September 4, 2020 | March 5, 2020 - April 2, 2020 | 84 | 376 | -Maximum likelihood phylogeny<br>-Bayiesian maximum clade credibility phylogeny<br>-Markov chain Monte Carlo Phylodynamics | [98] |
| Colombia<br>(South America) | September 6, 2020 | January 1, 2020 - April 19, 2020 | 43 | 1583 | -Maximum likelihood phylogeny | [183] |
| Karnataka (India) | July 24, 2020 | March 5 2020 - May 21, 2020 | 47 | 47 | Maximum likelihood phylogeny | [184] |
| Russia | July 17, 2020 | March 11 2020 - April 23 2020 | 135 | 19834 | Maximum likelihood phylogeny | [185] |
| Scotland | December 21, 2020 | February 2020 - March 2020 | 1,314 | 19370 | Maximum likelihood phylogeny | [186] |
| Austria | December 9th, 2020 | February 24 2020 - May 7 | 345 | 7666 | Maximum likelihood phylogeny | [187] |
| | | 2020 | | | | |
| Global | August 19th, 2020 | Up to March 30th, 2020 | 2492 | 2492 (all from GISAID database) | Amino-acid heterogeneity analysis | [188] |
| Ghana | December 16th, 2020 | March 12th–April 1st 2020, May 25th - 27th 2020 | 46 | 46 | Maximum likelihood phylogeny | [189] |
| Global | December 22nd, 2020 | April 17th, 2003 - March 24th, 2020 | 120 | 120 | Maximum likelihood phylogeny & MRP supertree method | [190] |
| Global | September 23rd, 2020 | December 2019 - April 18th 2020 | Greater than 6,000 | Greater than 6,000 | Maximum likelihood phylogeny | [191] |
| Lombardy (Italy) | March 29th, 2020 | February 21st, 2020 | 3 | 160 (157 from GISAID) | Bayesian Markov Chain Monte Carlo method | [192] |
| Chile | March 29th, 2020 | March 3rd - March 5th, 2020 | 4 | 222 (218 from GISAID) | Maximum likelihood phylogeny | [193] |
| Lazio (Italy) | August 26th, 2020 | February 27th, 2020 - March 23rd, 2020 | 8 | 150 | Maximum likelihood phylogeny | [194] |
| Turkey | June 21st, 2020 | May 1st, 2020 | 30 | 15,277 | Maximum likelihood phylogeny | [195] |
| Global | July 17th, 2020 | December 24th, 2019-February 9th, 2020 | 112 | 112 | Maximum likelihood phylogeny | [196] |

**Supplementary Table 3: Summary of wastewater analyses of SARS-CoV-2**

| The region covered (country) | Date of publication | Date of sample collection | Number of samples taken from WWTP (waste water treatment plants) | Total number of samples used for analysis (combined with publicly available data) | Method of NAAT Analysis | Citation |
|---|---|---|---|---|---|---|
| Federal State of North Rhine-Westphalia (Germany) | January 20, 2021 | April 8, 2020 | 9 | 9 | One-Step RT-qPCR | [197] |
| Rome and Milan (Italy) | September 20, 2020 | February 3, 2020 - April 2, 2020 | 12 | 12 | RT-PCR Analysis:ORFlab, Novel, nested RT-PCR targeting spike region | [198] |
| | | | | | RT-qPCR targeting RdRP gene | |
| Australia | August 1, 2020 | February 24, March 20, March 26, March 28-30, April 4, 2020 | 3 | 3 | RT-qPCR | [126] |
| Paris (France) | May 6, 2020 | March 2020 | 3 | 3 | RT-qPCR | [122] |
| The Netherlands | May 20, 2020 | February, March 2020 | 4 | 4 | RT-qPCR | [125] |
| New Haven (Connecticut, US) | September 18, 2020 | March 19,2020 - June 1, 2020 | 73 | 73 | qRT-PCR | [114] |
| Massachusetts (US) | April 7, 2020 | January 8, 2020<br>January 11, 2020<br>March 18-25, 2020 | 12 | 12 | RT-qPCR | [123] |
| Istanbul (Turkey) | May 6, 2020 | April 21-25, 2020 | 9 (7 WWTP and 2 manholes) | 9 | RT-qPCR | [199] |
| Israel | May 1, 2020 | N/A | 5 | 5 | qPCR | [200] |
| Valencia (Spain) | April 29, 2020 | February 12- April 14, 2020 | 15 | 15 | RT-qPCR | [121] |
| Wuhan (China) *used wastewater in septic tanks in hospitals | May 14, 2020 | February 26, 2020<br>March 1, 2020<br>March 10, 2020 | 3 | 3 | RT-qPCR | [201] |
| Santa Catalina (Brazil) | June 29, 2020 | October 30, 2019 - March 4, 2020 | 6 | 6 | RT-qPCR | [202] |
| Japan | June 18, 2020 | March 17, 2020 - May 7, 2020 | 13 (5 from WWTP and 3 from river) | 13 | qPCR<br>Nested PCR | [203] |
| India | June 18, 2020 | May 8, 2020<br>May 27, 2020 | N/A | N/A | RT-PCR | [204] |
| Bozeman (Montana) | September 22, 2020 | March 30, 2020 - June 12, 2020 | 17 | 17 | RT-PCR | [205] |
| Czech Republic | July 30, 2020 | April - June 2020 | 112 (from 33 different WWTP) | 112 | RT-qPCR | [206] |
| Alameda County<br>Marin County<br>(Northern California, US) | September 14, 2020 | May 19, 2020 - July 15, 2020 | N/A | N/A | qRT-PCR | [120] |
| Santa Clara County<br>(California, US) | September 15, 2020 | March 16 - May 31, June 2 - July 12 | 89 | 89 | RT-qPCR | [207] |
| Frankfurt (Germany) | October 27th, 2020 | April 2020 - August 2020 | 44 | 44 | RT-qPCR | [208] |
| Ottawa,Ontario and Gatineau, Quebec, Canada | October 23rd, 2020 | April 1st, 2020 - June 30th, 2020 | 23 (14 from Ottawa, and 9 from Gatineau) | 23 | RT-qPCR and RT-ddPCR | [209] |
| Istanbul(Turkey) | May 16th, 2020 | May 7th, 2020 | 9 | 9 | RT-qPCR | [210] |
| Quito River | November 15th, 2020 | June 5th, 2020 | 3 | 3 | qRT-PCR | [211] |
| (Ecuador) | | | | | | |
| Cruiseship in Australia, Flights from Los Angeles–Brisbane,Hong Kong–Brisban,. New Delhi–Sydney | July 5th, 2020 | April 23rd, 2020; July 6th, 2020; October 5th, 2020 | 21 | 21 | RT-qPCR and RT-ddPCR | [212] |
| Murcia (Spain) | 16 May 2020 | 2 March to 14 April,2020 | 72 | 72 | RT-qPCR | [213] |
| Milan Metropolitan Area, Italy | May 5, 2020. | April, 14th and April, 22th, 2020 | 16 | 16 | RT-PCR | [214] |
| Milan/Lombardy, Turin/Piedmont and Bologna/Emilia Romagna (Italy) | June 26, 2020 | 12 September 2018 and 19 June 2019, 9 October 2019 and 28 February 2020 | 40 | 40 | nested RT-PCR and RT-qPCR | [215] |
| Montpellier, France | July 09, 2020.July 09, 2020 | May 7th, 18th, 26th, June 4th, 15th and 25th | N/A | N/A | RT-qPCR | [216] |
| Louisiana, USA | November 15th, 2020 | January to April 2020 | 15 | 15 | RT-qPCR | [217] |
| Syracuse, NY and Onondaga County, NY | May 23rd, 2020 | May 6th and 13th, 2020 | 11 | 11 | RT-qPCR | [218] |